\documentclass[useAMS,usenatbib,usegraphicx]{mn2e}
\usepackage{epsf}
\usepackage{longtable}

\newcommand{\simgt}{\lower.5ex\hbox{$\; \buildrel > \over \sim \;$}}
\newcommand{\simlt}{\lower.5ex\hbox{$\; \buildrel < \over \sim \;$}}

\title[Gravitational lensing of gravitational waves]
{Effect of gravitational lensing on the distribution of gravitational
  waves from distant binary black hole mergers} 

\author[M.~Oguri]
{Masamune Oguri$^{1,2,3}$\thanks{E-mail: masamune.oguri@ipmu.jp} 
\\
$^1$Research Center for the Early Universe, University of Tokyo, 
7-3-1 Hongo, Bunkyo-ku, Tokyo 113-0033, Japan\\
$^2$Department of Physics, University of Tokyo, 7-3-1 Hongo,
Bunkyo-ku, Tokyo 113-0033, Japan\\
$^3$Kavli Institute for the Physics and Mathematics of the Universe
(Kavli IPMU, WPI), University of Tokyo, Chiba 277-8583, Japan\\
}

\begin{document}

\date{\today}

\voffset- .5in

\pagerange{\pageref{firstpage}--\pageref{lastpage}} \pubyear{}

\maketitle

\label{firstpage}

\begin{abstract}
The detailed observation of the distribution of redshifts and chirp
masses of binary black hole mergers is expected to provide a clue to
their origin. In this paper, we develop a hybrid model of the
probability distribution function of gravitational lensing magnification
taking account of both strong and weak gravitational lensing, and use
it to study the effect of gravitational lensing magnification on the
distribution of gravitational waves from distant binary black hole
mergers detected in ongoing and future gravitational wave
observations. We find that the effect of gravitational lensing
magnification is significant at high ends of observed chirp mass
and redshift distributions. While a high mass tail in the observed
chirp mass distribution is produced by highly magnified gravitational
lensing events, we find that highly {\it demagnified} images of strong
lensing events produce a high redshift ($z_{\rm obs}\ga 15$) tail in
the observed redshift distribution, which can easily be observed in
the third-generation gravitational wave observatories. Such a
demagnified, apparently high redshift event is expected to be
accompanied by a magnified image that is observed typically
$10-100$~days before the demagnified image. For highly magnified
events that produce apparently very high chirp masses, we expect pairs
of events with similar magnifications with time delays typically less
than a day. This work suggests the critical importance of
gravitational lensing (de-)magnification on the  interpretation of
apparently very high mass or redshift gravitational wave events. 
\end{abstract}


\section{Introduction}

Recent discoveries of gravitational waves from binary black hole (BH)
mergers open the possibility of using gravitational waves to probe the
Universe \citep{abbott16a}. These discoveries reveal the
population of binary BHs with their masses of $\sim 30~M_\odot$, whose
origin is still unknown. While such massive BHs can in principle be
formed as remnants of massive metal-poor stars, it is not very clear
whether binaries of these BHs that can merge within the age of the
Universe are sufficiently formed to explain the observed merger rate
of binary BHs \citep{abbott16c}. There are possible
scenarios of binary formations, including the formation from isolated
massive binary stars
\citep[e.g.,][]{kinugawa14,belczynski16,stevenson17} and the dynamical
formation in sense stellar systems
\citep[e.g.,][]{rodriguez16,oleary16}. In addition, such binary BHs
might be explained by primordial black holes  
\citep[PBHs;][]{bird16,sasaki16,clesse17}.

The distribution of binary BH mergers provides a clue to the origin of
binary BHs. From gravitational wave observations alone, one can obtain
information on masses and spins. Accurate observations of the
distributions of these properties help distinguish several binary BH
formation scenarios \citep[e.g.,][]{farr17,kocsis18}. Another
information may be provided by the distribution of luminosity
distances, or redshifts inferred from the luminosity distances. For
instance, \citet{nakamura16} and \citet{koushiappas17} argue that the
distribution of binary BH mergers at very high redshifts is a powerful
discriminant of its origin.

However parameters derived from observations of binary BH mergers are
affected by gravitational lensing. For instance, the effect of weak
gravitational lensing on binary BH mergers has been considered in the
context of the so-called standard siren method to constrain the
distance-redshift relation
\citep[e.g.,][]{markovic93,holz05,bertacca18} and cross-correlation of
their spatial distributions with large-scale structure
\citep[e.g.,][]{camera13,namikawa16,oguri16,osato18}. In particular,
gravitational lensing magnification shifts the luminosity distance
estimated from the merger waveform, and therefore the redshift of the
merger event inferred from the luminosity distance is biased if the
lensing magnification is not corrected for. Since we can measure only
``redshifted'' BH masses from the merger waveform, BH masses from the
merger waveform can also be biased due to gravitational lensing
magnification. This indicates that the presence of highly magnified
events can produce a heavy high mass tail in the BH mass distribution
inferred from gravitation wave observations
\citep{dai17a,smith18,broadhurst18}.    

When the gravitational lensing effect is strong, we observe multiple
images of binary BH mergers. Depending on the frequency of
gravitational wave observations, the mass of a lensing object, and the
image configuration, the wave effect can play an important role
\citep[e.g.,][]{nakamura99,takahashi03,takahashi17,diego18}.
Predictions for the observed number of strongly lensed binary BH
mergers
\cite[e.g.,][]{sereno10,sereno11,biesiada14,ding15,liao17,ng18,li18} 
indicate that a large number of such events can be discovered in
next-generation gravitational wave observatories.

In this paper, we study the effect of gravitational lensing
magnification on the distribution of gravitational waves from binary
BH mergers. We focus on how the distribution of {\it observable} 
quantities, such as luminosity distances and BH masses inferred from
waveforms, is affected by gravitational lensing. In this regard, our
work is similar to that conducted by \citet{dai17a}, but in our work we
explore the role of strong gravitational lensing more thoroughly. Our
working assumption is that, due to the poor localization accuracy and
the difficulty in identifying electromagnetic counterparts of binary
BH mergers \citep[e.g.,][]{abbott16b}, the identification of multiple 
images is not straightforward in gravitational wave observations. We
develop a hybrid framework to compute the magnification
probability distribution function (PDF) for which we combine the
effects of both weak and strong gravitational lensing and treat
multiple images separately. We show that demagnified images play an
important role in the distribution of binary BH mergers at high
redshifts. We also discuss the prospect for identifying possible
multiple image pairs of gravitational wave events in future
observations. 

This paper is organized as follows. In Section~\ref{sec:gl_model}, we
describe our hybrid model of the magnification PDF. We present models
of binary BH mergers adopted in the paper in Section~\ref{sec:bbh}.
The method to compute distributions of binary BH mergers with
and without the gravitational lensing magnification in
Section~\ref{sec:lensdist}. We present our results in
Section~\ref{sec:results}, and summarize our results in
Section~\ref{sec:summary}. 
Throughout the paper, we assume a flat cosmological model with matter
density $\Omega_{\rm M}=0.3156$, baryon density $\Omega_{\rm b}=0.04917$, 
cosmological constant $\Omega_\Lambda=0.6727$, the dimensionless
Hubble constant $h=H_0/(100\,{\rm km\,s^{-1}Mpc^{-1}})=0.6727$, 
spectral index $n_{\rm s}=0.9645$, and the
normalization of density fluctuations $\sigma_8=0.831$ 
\citep{planck16}. 

\section{Model of gravitational lensing}
\label{sec:gl_model}

In this Section, we present our model of the magnification PDF for
compact astronomical sources. In order to take account of both strong
and weak lensing effects, we adopt a hybrid approach in which the
magnification PDF at low and high magnifications are computed
separately. Our model allows us to compute magnification PDFs for wide
ranges of source redshifts and magnifications, and to study overall
distributions of weakly and strongly lensed sources as well as strong
lensing properties for a subsample of sources in a unified manner. 

\subsection{Magnification PDF at low magnifications}
\label{sec:magpdf_low}

Cosmological PDFs of lensing magnifications have been studied using
various methods including ray-tracing in numerical simulations 
\citep[e.g.,][]{wambsganss97,hamana00,takada03,hilbert07,hilbert08,takahashi11,castro18} 
as well as analytical approaches
\citep[e.g.,][]{schneider88,holz98,perrotta02,wyithe02,yoo08,lima10,kainulainen11,lapi12,fialkov15}. 
These studies showed that the magnification PDF significantly deviates
from the Gaussian distribution such that it has a long tail toward
high magnifications. The behavior of the very high magnification tail
is dominated by strong lensing and is well understood by catastrophe
theory \citep[e.g.,][]{blandford86}.

In this paper, we compute the magnification PDF at low magnifications
following the methodology developed in \citet{takahashi11}. In this
method, the magnification PDF is computed from the convergence PDF
adopting an approximate relation
\begin{eqnarray}
\mu=\frac{1}{(1-\kappa)^2},
\label{eq:kappa2mu}
\end{eqnarray}
where $\mu$ and $\kappa$ denote magnification and convergence,
respectively. For the convergence PDF $dP/d\kappa$, we adopt a model
of \citet{das06} that is given by
\begin{eqnarray}
 \frac{dP}{d\kappa}&=&N_\kappa \exp \left[ -\frac{1}{2 \omega^2_\kappa}  
 \left\{ \ln \left( 1+ \frac{\kappa}{\left| \kappa_{\rm empty} \right|}
 \right) + \frac{\omega^2_\kappa}{2} \right\}^2 \right. \nonumber \\
 && \left. \times \left\{ 1+\frac{A_\kappa}{1+ \kappa / \left| \kappa_{\rm empty}
 \right|} \right\} \right] \frac{1}{\kappa+\left| \kappa_{\rm empty}
 \right|},
\end{eqnarray}
where parameters $N_\kappa$, $\omega_\kappa$, and $A_\kappa$ are determined
numerically so as to satisfy 
\begin{eqnarray}
\int d\kappa\frac{dP}{d\kappa}=1,
\end{eqnarray}
\begin{eqnarray}
\int d\kappa\frac{dP}{d\kappa}\kappa=-2\langle\kappa^2\rangle,
\end{eqnarray}
\begin{eqnarray}
\int d\kappa\frac{dP}{d\kappa}\kappa^2=\langle\kappa^2\rangle.
\end{eqnarray}
The discussion on the negative mean convergence is found in
\citet{takahashi11} and also in e.g., \citet{kaiser16}.
In this model, the information on the source redshift, the matter
power spectrum, and cosmological parameters is included in 
$\kappa_{\rm empty}$ and $\langle\kappa^2\rangle$. The former
denotes the minimum convergence value, which is realized when light
ray propagates through the empty region i.e., the density fluctuation
$\delta=-1$. Specifically $\kappa_{\rm empty}$ for a source at
$z=z_{\rm s}$ is computed as
\begin{equation}
  \kappa_{\rm empty}=
-\frac{3\Omega_{\rm M}}{2} \left(\frac{H_0}{c}\right)^2
\int_0^{z_{\rm s}}
 \frac{c\,dz}{H(z)} \left( 1+z \right) \frac{r(\chi)r(\chi_{\rm
     s}-\chi)}{r(\chi_{\rm s})},
\label{eq:kappa_empty}
\end{equation}
where $H(z)$ is the Hubble parameter at redshift $z$, $\chi$ is the
radial distance $\chi(z)=\int_0^z c\,dz'/H(z')$, and $r(\chi)$ is the
comoving angular diameter distance that is equal to $\chi$ for a flat
universe. On the other hand, the variance of the convergence
$\langle\kappa^2\rangle$ is related with the matter power spectrum
$P_{\rm m}(k,z)$ as
\begin{eqnarray}
 \langle \kappa^2 \rangle & = & \frac{9\Omega_{\rm M}^2}{8 \pi}
\left(\frac{H_0}{c}\right)^4 \int_0^{z_{\rm s}}
 \frac{c\,dz}{H(z)} \left( 1+z \right)^2 
 \nonumber \\
&&\times \left[ \frac{r(\chi)r(\chi_{\rm
     s}-\chi)}{r(\chi_{\rm s})}\right]^2\int dk \,k \,P_{\rm m}(k,z).
\label{eq:kappa_sig}
\end{eqnarray}
We compute the nonlinear matter power spectrum using the improved {\tt
  halofit} model of \citet{takahashi12}.

The convergence PDF is then converted to the magnification PDF using
equation~(\ref{eq:kappa2mu})
\begin{equation}
 \frac{dP_{\kappa}}{d\mu}=
\left.\frac{(1-\kappa)^3}{2}\frac{dP}{d\kappa}\right|_{\kappa=1-1/\sqrt{\mu}}.
\label{eq:dpkdmu}
\end{equation}
\citet{takahashi11} showed that the magnification PDFs derived above
fit those from ray-tracing simulations quite well up to $\mu\sim 3$.
Since we compute the magnification PDF at the high-magnification
region separately based on strong lensing statistics
(Section~\ref{sec:magpdf_high}), we explicitly truncate the
magnification PDF derived in equation~(\ref{eq:dpkdmu}) as
\begin{equation}
 \frac{dP_{\rm wl}}{d\mu}=\frac{dP_{\kappa}}{d\mu}\exp
\left[-\left(\frac{\mu}{\mu_0}\right)^4\right],
\label{eq:dpwldmu}
\end{equation}
where we adopt $\mu_0=3$.

We note that we do not take account of any baryonic effect here. 
Specifically, the matter power spectrum used in equation~(\ref{eq:kappa_sig}) 
is the one derived from dark matter only $N$-body simulations. 
This is reasonable because the main effect of baryon physics on the
magnification PDF is to enhance the high magnification tail at
$\mu\ga 3$, and the effect of baryon on the magnification PDF around
the peak ($\mu\sim 1$) appears to be insignificant 
\citep[see e.g.,][]{hilbert08,castro18}. In
Section~\ref{sec:magpdf_high}, we derive the magnification PDF at high
magnifications taking proper account of baryon effects. 

\subsection{Magnification PDF at high magnifications}
\label{sec:magpdf_high}

At high magnifications, the PDF is dominated by strong lensing
events, for which the effect of baryon cooling is significant
\citep[e.g.,][]{kochanek01}. We follow the Monte-Carlo method
developed in \citet{oguri10b} to derive strong lens properties of
distant compact sources, which we use to derive the magnification
PDF at high magnifications. In short, we randomly generate a sample of
galaxies follow the velocity dispersion function of galaxies, and
randomly assign ellipticities and external shear for each galaxy, and 
solve the lens equation using the {\tt glafic} code \citep{oguri10a}
to check whether randomly generated sources are multiply imaged or not
\citep[see][for more details]{oguri10b}. We use the mock strong lens
samples generated by this method to construct the magnification PDF
expected from strong lensing events.

\begin{figure}
\begin{center}
 \includegraphics[width=1.0\hsize]{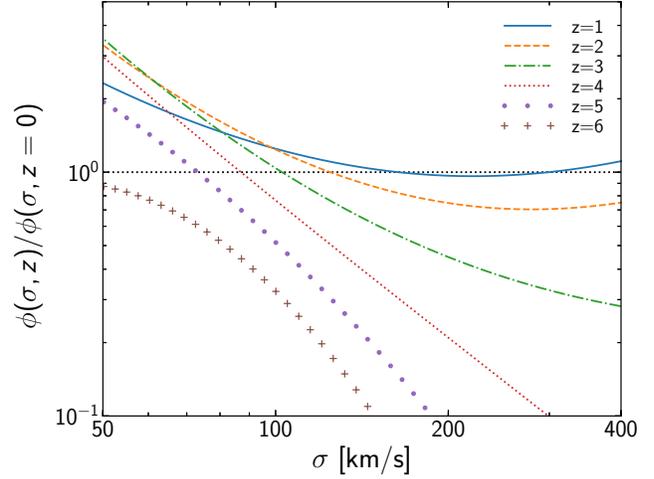}
\end{center}
\caption{The redshift evolution of the velocity dispersion function
  adopted in this paper (equation~\ref{eq:vf_evolution}). We show the
  velocity functions at several different redshifts relative to the
  local velocity function at $z=0$ as a function of the velocity
  dispersion $\sigma$.  
\label{fig:vf}}
\end{figure}

While we mostly follow the methodology developed in \citet{oguri10b},
here we include several updates of calculations. First, in this paper
we use an updated velocity dispersion function of all-type galaxies
derived from the Sloan Digital Sky Survey Data Release 6
\citep{bernardi10} 
\begin{equation}
\phi_{\rm loc}(\sigma)=\phi_*
\left(\frac{\sigma}{\sigma_*}\right)^\alpha
\exp\left[-\left(\frac{\sigma}{\sigma_*}\right)^\beta\right]
\frac{\beta}{\Gamma(\alpha/\beta)}\frac{1}{\sigma},
\end{equation}
with $\phi_*=2.099\times 10^{-2}(h/0.7)^3{\rm Mpc^{-3}}$,
$\sigma_*=113.78~{\rm km\,s^{-1}}$, $\alpha=0.94$, and $\beta=1.85$.
We note that \citet{oguri10b} adopted a velocity function of
early type galaxies derived by \citet{choi07}. Different choices of a
velocity function result in nearly a factor of two difference in total
strong lensing probabilities, and therefore the same factor of
difference in the magnification PDF at high magnifications. 
Since we are interested in strong lensing of very high-redshift
sources, redshifts of lensing galaxies can also be relatively high.
In \citet{oguri10b}, the velocity dispersion function is assumed to
not evolve with redshift, which may be a reasonable assumption out to
$z\sim 1.5$ \citep[e.g.][]{bezanson11}. In this paper, however, we are
interested in strong lensing of very high redshift sources for which
redshifts of lensing galaxies can be much higher than $z\sim 1.5$. In
order to take account of the redshift evolution of the velocity
dispersion function,  we adopt the result of \citet{torrey15} in which
the redshift evolution of velocity dispersion functions is derived
from the Illustris cosmological hydrodynamical simulation. \citet{torrey15}
provided a velocity dispersion function $\phi_{\rm hyd}(\sigma,z)$ that
fits results of the hydrodynamical simulation from $z=0$ to $6$. 
We combine this redshift dependence of the velocity dispersion
function with the accurate local velocity dispersion function
measurement by \citet{bernardi10} to derive the velocity dispersion
function that is applicable for a wide range of redshift as 
\begin{equation}
\phi(\sigma,z)=\phi_{\rm loc}(\sigma)
\frac{\phi_{\rm hyd}(\sigma,z)}{\phi_{\rm hyd}(\sigma,0)}.
\label{eq:vf_evolution}
\end{equation}
We show the redshift evolution of the velocity dispersion function
derived from equation~(\ref{eq:vf_evolution}) in Figure~\ref{fig:vf}.
Since the strong lensing probability is proportional to $\sigma^4$,
velocity dispersions of typical strong lensing galaxies correspond to
the peak of $\sigma^4\phi(\sigma)$ and hence those of massive galaxies
with high velocity dispersions. For these massive, high velocity
dispersion galaxies, the redshift evolution is weak up to $z\sim 1$,
but shows the strong decline of the number density at $z\ga 2$, which
is in line with a native expectation from the redshift evolution of
the mass function of dark matter haloes.

The density profile of each galaxy is same as that adopted in
\citet{oguri10b}. We adopt the singular isothermal ellipsoid model for
the galaxy, which is known to approximate the density profile of early
type galaxies well, and add external shear. The probability 
distributions of the ellipticity and external shear are assumed to be
same as those used in \citet{oguri10b}, i.e., the Gaussian
distribution with mean of 0.3 and dispersion of 0.16 for the
ellipticity and the lognormal distribution with mean 0.05 and
dispersion 0.2~dex for the external shear. Their position angles are
assumed to be completely random. In this paper, we ignore the
dynamical normalization of the singular isothermal ellipsoid model for
simplicity \citep[see][]{oguri10b}. 

We consider only galaxies as lensing objects, because total strong
lensing cross sections for compact sources are dominated by those of
single massive galaxies \citep[e.g.,][]{keeton05}, which is also
supported by the statistics of strong gravitational lenses discovered in
submillimetre surveys \citep[e.g.,][]{amvrosiadis18}. This means 
that the inclusion of group- and cluster-scale haloes in the
computation does not change our quantitative results significantly. 

From the mock strong lens sample, we construct the source plane
magnification PDF as a function of redshift, which we denote $dP_{\rm
  sl}/d\mu$. In the case of strong lensing, there is an ambiguity
about how to deal with multiple images. In this paper, we regard
multiple images as distinct images in computing the magnification PDF
from the mock lens sample, because in observations of binary BH
mergers it is not straightforward to identify multiple images. See 
Appendix~\ref{sec:magpdf_def} for more detailed discussions about the
definition of magnification PDFs in the presence of multiple images. 

\begin{figure}
\begin{center}
 \includegraphics[width=1.0\hsize]{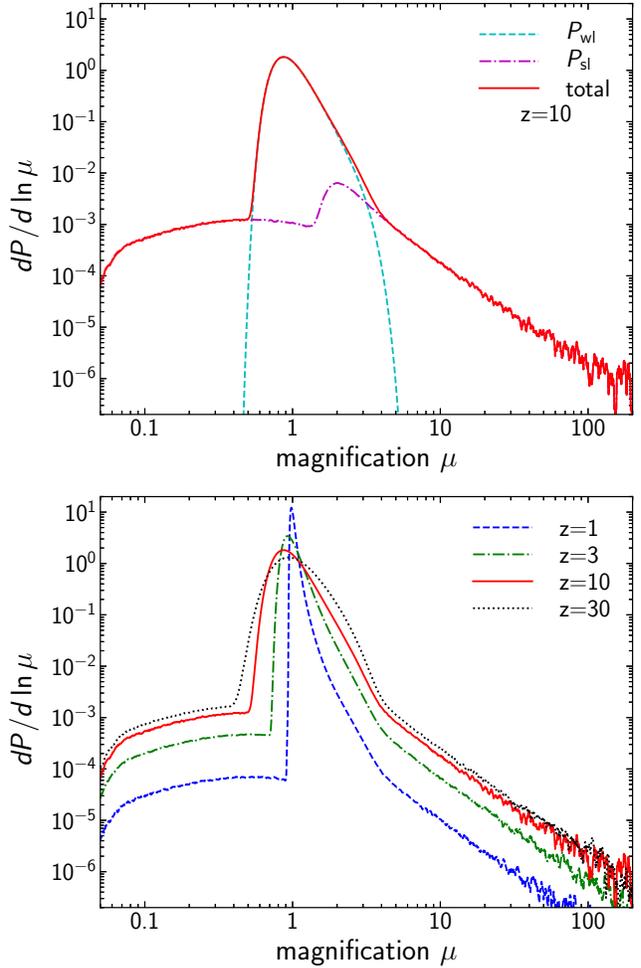}
\end{center}
\caption{{\it Top:} The total magnification PDF for $z=10$ ({\it
    solid}) is shown together with contributions from magnification
  PDFs at low and high magnifications derived in
  Sections~\ref{sec:magpdf_low} and \ref{sec:magpdf_high},
  respectively. These magnification PDFs derived using different
  methods are combined based on the methodology detailed in
  Appendix~\ref{sec:magpdf_def}, in which multiple images are treated
  separately. 
  {\it Bottom:} The magnification PDFs for $z=1$ ({\it dashed}), $3$
  ({\it dash-dotted}), $10$ ({\it solid}), and $30$ ({\it dotted}).
\label{fig:mag_pdf}}
\end{figure}

\begin{figure}
\begin{center}
 \includegraphics[width=1.0\hsize]{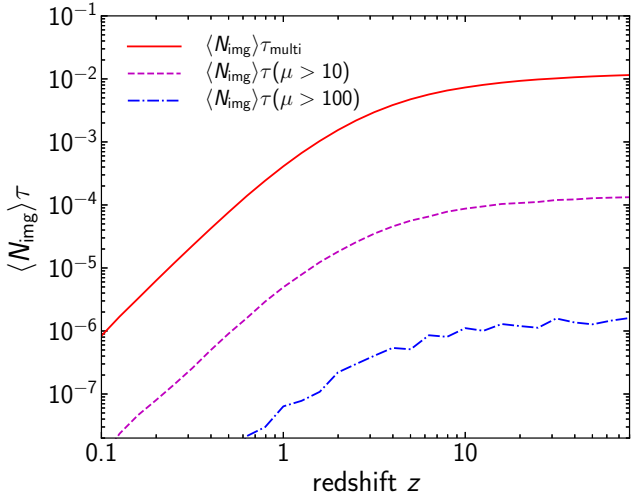}
\end{center}
\caption{Optical depths $\langle N_{\rm img}\rangle\tau$ are plotted as a
  function of redshift. Here we define $\langle N_{\rm img}\rangle\tau$
  regarding multiple images as distinct images (see text for more
  details). The solid line shows the optical depth for multiple images
  (equation~\ref{eq:tau_multi_n}), whereas the dashed and
  dash-dotted lines show the optical depths defined in
  equation~(\ref{eq:tau_mu}) with $\mu_{\rm th}=10$ and $100$,
  respectively. 
\label{fig:tau}}
\end{figure}

\subsection{Combined magnification PDF}
\label{sec:magpdf_comb}

We follow the procedure detailed in Appendix~\ref{sec:magpdf_def} to
combine magnification PDFs at low (Section~\ref{sec:magpdf_low}) and
high (Section~\ref{sec:magpdf_high}) magnifications. As described in
Appendix~\ref{sec:magpdf_def}, we include minor corrections in order
to ensure the correct normalization and the mean magnification. The
magnification PDFs are computed as a function of source redshift
$z_{\rm s}$ with the source redshift bin size of 0.1~dex.

We show examples of magnification PDFs in Figure~\ref{fig:mag_pdf}. As
discussed in Appendix~\ref{sec:magpdf_def}, since we treat multiple
images separately, the normalization of the magnification PDF exceeds
unity, though only slightly. With increasing redshift, the
magnification PDF becomes wide and has a higher tail at high
magnifications. The feature seen at $\mu\sim 2$ for the magnification
PDF from strong lensing, $P_{\rm sl}$ corresponds to the transition
between magnified and demagnified images.

From the magnification PDF, we can compute the optical depth $\tau$
for strong lensing, which represents the probability of a source at
redshift $z$ being strongly lensed. Again, since we treat multiple
images separately, this optical depth differs from the conventional
definition of the optical depth for which multiple images are grouped
together. First we consider the optical depth for all multiple images,
which was defined in equation~(\ref{eq:tau_multi_n}). We also define
the optical depth defined by the magnification threshold as
\begin{equation}
  \langle N_{\rm img}\rangle\tau(>\mu_{\rm th})=
  \int^\infty_{\mu_{\rm th}}d\mu\frac{dP}{d\mu},
  \label{eq:tau_mu}
\end{equation}
where we use the total magnification PDF for $dP/d\mu$, although at
high magnifications it is dominated by that from strong lens mocks
derived in Section~\ref{sec:magpdf_high}. In Figure~\ref{fig:tau},
we show optical depths with different definitions as a function of
redshift. In all cases shown in the Figure, the optical depth rapidly
increases as a function of redshift out to $z\sim 3$, but at high
redshifts $z\ga 10$ the redshift dependence is rather weak.

\section{Models of black hole binaries}
\label{sec:bbh}

\begin{table}
 \caption{Parameters for BH merger rate density
   (equation~\ref{eq:r_gw}) and the chirp mass distribution
   (equation~\ref{eq:mc_dist}) for three stellar origin models,
   Pop-I/II \citep{belczynski17}, Pop-III (B17) \citep{belczynski17},
   and Pop-III (K16) \citep{kinugawa16}. \label{tab:bhdist}}    
 \begin{tabular}{@{}cccc}
 \hline
   Parameters
   & Pop-I/II
   & Pop-III (B17)
   & Pop-III (K16)
   \\
 \hline
$a_1$ & $6.6\times 10^3$ & $6\times 10^4$ & $1\times 10^4$ \\
$a_2$ & 1.6 & 1.0 & 0.7 \\
$a_3$ & 2.1 & 1.4 & 1.1 \\
$a_4$ & 30  & $3\times 10^6$ & 500 \\
$z_{\rm trunc}$ &  15 & 45 & 45 \\
$b_1$ & 8  & 28 & 20 \\
$b_2$ & 30 & 30 & 72 \\
 \hline
 \end{tabular}
\end{table}

\subsection{Stellar Origins}
\label{sec:bbh_stellar}

It has been known that BHs form naturally from the collapse of massive
stars at the final stage of their evolution. This suggests that binary
BHs may form from massive binary stars. Because the evolution of
massive stars is sensitive to the metallicity due to its large impact
on the mass loss rate, the merger rate density and masses of binary
BHs are also expected to be sensitive to the metallicity. In addition,
uncertainties associated with initial binary parameters make the
prediction on the BH merger rate density quite uncertain
\citep[e.g.,][]{belczynski02,belczynski10,belczynski16,belczynski17,dominik12,dominik13,dominik15,kinugawa14,kinugawa16,hartwig16}. 

In this paper, we consider a few model predictions from the population
synthesis calculations as representative examples. Since the merger
rate density and BH mass distribution depend sensitively on
metallicity, those of metal free Population III (Pop-III) stars are
expected to be markedly different from Population I and II (Pop-I/II)
stars. For Pop-I/II stars, we adopt the model ``M10'' presented in
\citet{belczynski17}. For Pop-III stars, we consider a model presented
in \citet{kinugawa16} (model ``Standard'') and \citet{belczynski17}
(``FS1'') to cover possible ranges of model predictions. 

While in \citet{belczynski17} and \citet{kinugawa16} the redshift and
mass distributions of BH mergers have been derived numerically
with the binary population synthesis models, in this paper we adopt 
simple analytic forms for these distributions that roughly reproduce
their numerical results. For the BH merger rate density, we assume the
following functional form 
\begin{equation}
  \frac{R_{\rm GW}(z)}{\rm Gpc^{-3}yr^{-1}}=\frac{a_1e^{a_2z}}{e^{a_3z}+a_4},
\label{eq:r_gw}
\end{equation}
for $z<z_{\rm trunc}$ and $R_{\rm GW}(z)=0$ at $z>z_{\rm trunc}$. On
the other hand, we assume the following form for the chirp mass
distribution of BH mergers
\begin{equation}
  \frac{dp}{d\mathcal{M}}\propto \mathcal{M}^{-2.3}
\left[1-e^{-(\mathcal{M}/b_1)^8}\right]e^{-(\mathcal{M}/b_2)^8}.
\label{eq:mc_dist}
\end{equation}
Note that the normalization is determined so as to satisfy $\int
(dp/d\mathcal{M}) d\mathcal{M}=1$. Parameters for these distributions
are summarized in Table~\ref{tab:bhdist}. 

\begin{figure}
\begin{center}
 \includegraphics[width=1.0\hsize]{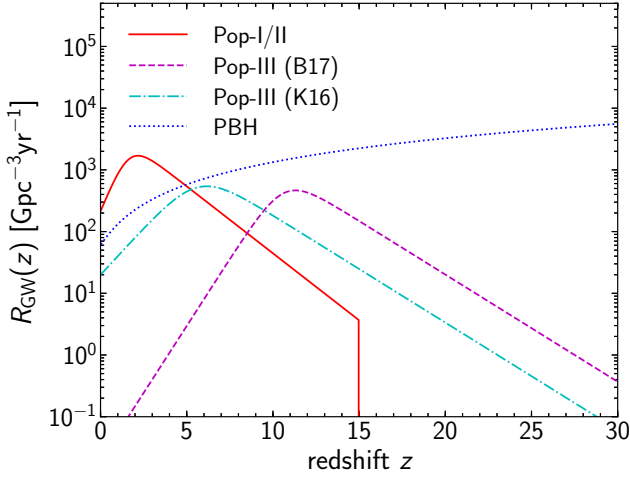}
\end{center}
\caption{BH merger rate densities as a function of redshift
  for the three stellar origin models described in
  Section~\ref{sec:bbh_stellar} and the PBH model described in
  Section~\ref{sec:bbh_pbh}. 
\label{fig:mrate}}
\end{figure}

\begin{figure}
\begin{center}
 \includegraphics[width=1.0\hsize]{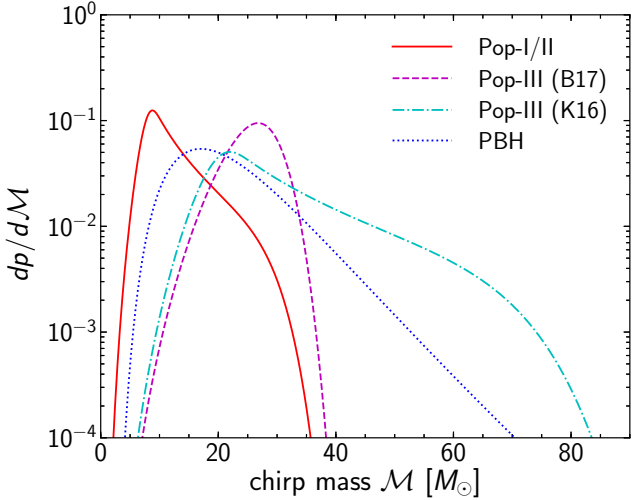}
\end{center}
\caption{Distributions of the chirp mass for the three stellar origin
  models described in Section~\ref{sec:bbh_stellar} and the PBH model
  described in Section~\ref{sec:bbh_pbh}. 
\label{fig:dist_bhm}}
\end{figure}

We show BH merger rate densities and chirp mass distributions for
these three models in Figures~\ref{fig:mrate} and \ref{fig:dist_bhm},
respectively. 

\subsection{Primordial black holes}
\label{sec:bbh_pbh}

Another scenario is that observed BH mergers may originate from
PBHs \citep[see][for a review]{sasaki18}.
In this paper, we consider a binary formation model considered in
\citet{sasaki16}, which originated from work by \citet{nakamura97}.
In this scenario, PBHs created in the early universe form a binary
with high eccentricity due to the tidal effect of a neighboring PBH. 
Here we simply adopt the expression of the BH merger rate density as a
function of redshift presented in \citet{sasaki16}. We set the mass 
fraction of PBH, $f_{\rm PBH}=\Omega_{\rm PBH}/\Omega_{\rm DM}$, to
$f_{\rm PBH}=5\times 10^{-3}$ so that it roughly matches the observed
BH merger rate density \citep{abbott16c}.

Although a single PBH mass has been considered in \citet{sasaki16}, in
this paper we include the mass distribution of PBHs by assuming a
log-normal form with the median chirp mass of $20~M_\odot$ and the
scatter of $\ln \mathcal{M}$ of 0.4. The merger rate density and chirp
mass distribution of the PBH model are compared with stellar origin
models in Figures~\ref{fig:mrate} and \ref{fig:dist_bhm}, respectively. 

\section{Calculation of the Distribution of binary black hole mergers}
\label{sec:lensdist}

\subsection{Signal-to-noise ratio}
\label{sec:sn_calc}

In this paper, we consider only the inspiral phase of gravitational
waves to compute the expected signal-to-noise ratio for
simplicity. This assumption has also been used in the literature to
discuss detectabilities of binary BH mergers in future
detectors \citep[e.g.,][]{taylor12,miyamoto17,li18}. In this case, the
signal-to-noise ratio $\rho$ of binary BH mergers with masses
$m_1$ and $m_2$ is computed as \citep{finn96}
\begin{equation}
  \rho=\sqrt{\frac{5}{96\pi^{4/3}}}\frac{R_\odot}{D_{\rm L}(z)}
  \left(\frac{\mathcal{M}_z}{M_\odot}\right)^{5/6}\Theta\sqrt{I}
  \equiv \rho_0\Theta,
  \label{eq:rho_def}
\end{equation}
\begin{equation}
R_\odot=cT_\odot =\frac{GM_\odot}{c^2},
\end{equation}
\begin{equation}
\mathcal{M}_z=(1+z)\mathcal{M}=(1+z)\frac{(m_1m_2)^{3/5}}{(m_1+m_2)^{1/5}},
\end{equation}
\begin{equation}
I=\int_0^{f_{\rm max}} df\,T_\odot^{-1/3}f^{-7/3}\{S_n(f)\}^{-1},
\end{equation}
where $D_{\rm L}(z)$ is the luminosity distance, $\mathcal{M}_z$ is the
redshifted chirp mass, and $S_n(f)$ is the noise power spectrum
density of a detector which has the dimension of Hz$^{-1/2}$.  The
angular orientation function $\Theta$ encapsulates information on the
detector with respect to the position of the binary BH merger on the
sky as well as the inclination angle of the merger event. Assuming the
random orientations, the PDF of $\Theta$ can be well approximated by
\citep{finn96} 
\begin{equation}
  P(\Theta)=\frac{5\Theta(4-\Theta)^3}{256},
  \label{eq:theta_pdf}
\end{equation}
for $0<\Theta<4$ and $P(\Theta)=0$ otherwise.
We assume that $f_{\rm max}$ corresponds to the
frequency at the innermost stable circular orbit (ISCO) that is given
by
\begin{equation}
f_{\rm ISCO}=\frac{M_\odot}{6^{3/2}\pi T_\odot
  (1+z)M}\approx\frac{4397\,{\rm Hz}}{(1+z)(M/M_\odot)},
\end{equation}
where $M=m_1+m_2$ is the total mass of the binary BH system. 
For simplicity, throughout the paper we assume that masses of binary
BHs are always equal e.g., $M=2^{6/5}\mathcal{M}$, to compute $f_{\rm
  ISCO}$.

\subsection{Distribution of binary BH mergers}
\label{sec:gwdist_obs}

First we derive the event rate of binary BH mergers for a given
gravitational wave observatory without the effect of gravitational
lensing magnification. Assuming a threshold of the signal-to-noise
ratio of $\rho_{\rm th}$, the event rate $R_{\rm obs}$ is computed as
\begin{equation}
R_{\rm obs}=\int dz\int d\mathcal{M} \frac{dV}{dz}\frac{R_{\rm
    GW}(z)}{1+z}\frac{dp}{d\mathcal{M}}S(\rho_{\rm th}; \mathcal{M}, z),
\label{eq:robs_nolens}
\end{equation}
where $R_{\rm GW}(z)$ and $dp/d\mathcal{M}$ are the BH merger rate
density and the chirp mass distribution, respectively, presented in 
Section~\ref{sec:lensdist}, $dV/dz$ is the comoving volume element,
and a factor $1/(1+z)$ takes account of the cosmological time
dilation. The effect of the signal-to-noise ratio threshold $\rho_{\rm
  th}$ is included in $S(\rho_{\rm th}; \mathcal{M}, z)$ as
\begin{equation}
S(\rho_{\rm th}; \mathcal{M}, z)=T(4)-T(\rho_{\rm th}/\rho_0),
\end{equation}
\begin{equation}
T(\Theta)=\frac{\Theta^2}{256}(160-80\Theta+15\Theta^2-\Theta^3),
\end{equation}
for $\rho_{\rm th}/\rho_0<4$ and $S(\rho_{\rm th}; \mathcal{M}, z)=0$ otherwise. 

Next we consider the effect of gravitational lensing magnification. 
Ignoring the effect of the phase shift \citep{dai17b}, we can include
the effect of lensing magnification $\mu$ in
the geometric optics limit simply by shifting the luminosity distance
as  
\begin{equation}
  D_{\rm L}(z) \rightarrow \frac{D_{\rm L}(z)}{\sqrt{\mu}}.
  \label{eq:dis_lum_mag}
\end{equation}
Therefore, in presence of the lensing effect, the event rate
is computed as 
\begin{eqnarray}
R_{\rm obs}&=&\int dz \int d\mu \frac{dP}{d\mu}\int d\mathcal{M} \frac{dV}{dz}\frac{R_{\rm
    GW}(z)}{1+z}\frac{dp}{d\mathcal{M}}\nonumber\\
&&\times S_{\rm lens}(\rho_{\rm th}; \mathcal{M}, z, \mu),
\label{eq:robs}
\end{eqnarray}
\begin{equation}
S_{\rm lens}(\rho_{\rm th}; \mathcal{M}, z, \mu)=T(4)-T(\rho_{\rm th}/(\sqrt{\mu}\rho_0)),
\end{equation}
for $\rho_{\rm th}/(\sqrt{\mu}\rho_0)<4$ and $S_{\rm lens}(\rho_{\rm
  th}; \mathcal{M}, z, \mu)=0$ otherwise, and $dP/d\mu$ is the
magnification PDF as a function of redshift derived in
Section~\ref{sec:gl_model}. 

\begin{figure}
\begin{center}
 \includegraphics[width=1.0\hsize]{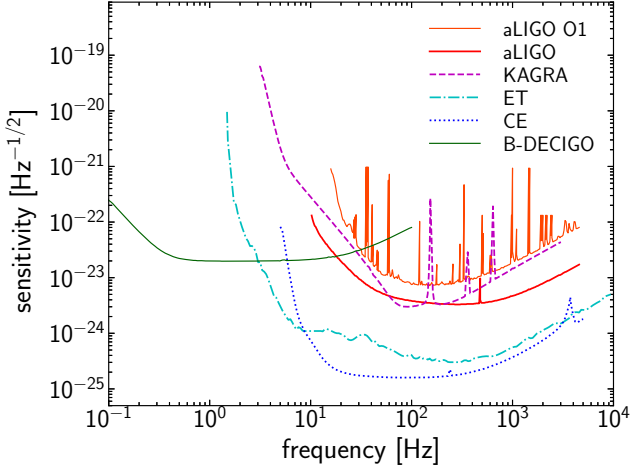}
\end{center}
\caption{Noise power spectra $S_n(f)$ for various gravitational wave
  observatories that are considered in this paper. The sensitivity of
  advanced LIGO O1 is also shown for reference.
\label{fig:sens}}
\end{figure}

In this paper, we consider how gravitational lensing modifies the
{\it observable} distribution of BH mergers. Specifically, we consider
the differential distributions of the ``observed redshift'' $z_{\rm
  obs}$, which is the redshift inferred from the luminosity distance
without the correction of lensing magnification $\mu$, as well as
the ``observed chirp mass'' $\mathcal{M}_{\rm obs}$, which is the
chirp mass inferred from the observed waveform, again without the
correction of lensing magnification. They are simply defined as 
\begin{equation}
  D_{\rm L}(z_{\rm obs})=\frac{D_{\rm L}(z)}{\sqrt{\mu}},
  \label{eq:z_obs}
\end{equation}
\begin{equation}
  \mathcal{M}_{\rm obs}=\frac{1+z}{1+z_{\rm obs}}\mathcal{M}.
  \label{eq:m_obs}
\end{equation}
By differentiating equation~(\ref{eq:robs}) we can obtain differential
distribution of the event rate, $dR_{\rm obs}/dz_{\rm obs}$
and $dR_{\rm obs}/d\mathcal{M}_{\rm obs}$.

\subsection{Gravitational wave observatories}
\label{sec:exps}

In our calculation, information on gravitational wave observatories is
included in the noise power spectrum $S_n(f)$. As specific examples,
we consider $S_n(f)$ from ongoing observatories such as advanced LIGO
(aLIGO)\footnote{https://www.ligo.caltech.edu} for the design
specification and KAGRA \citep{nakamura16}, as well as the so-called
third generation observatories such as Einstein Telescope
\citep[ET;][]{regimbau12}  and Cosmic Explorer \citep[CE;][]{CE17}. 
We also consider a planned space mission B-DECIGO \citep{nakamura16}
which is supposed to find binary BH mergers out to high redshifts.
The noise power spectra assumed in this paper are shown in
Figure~\ref{fig:sens}. 
 
\begin{figure*}
\begin{center}
 \includegraphics[width=0.95\hsize]{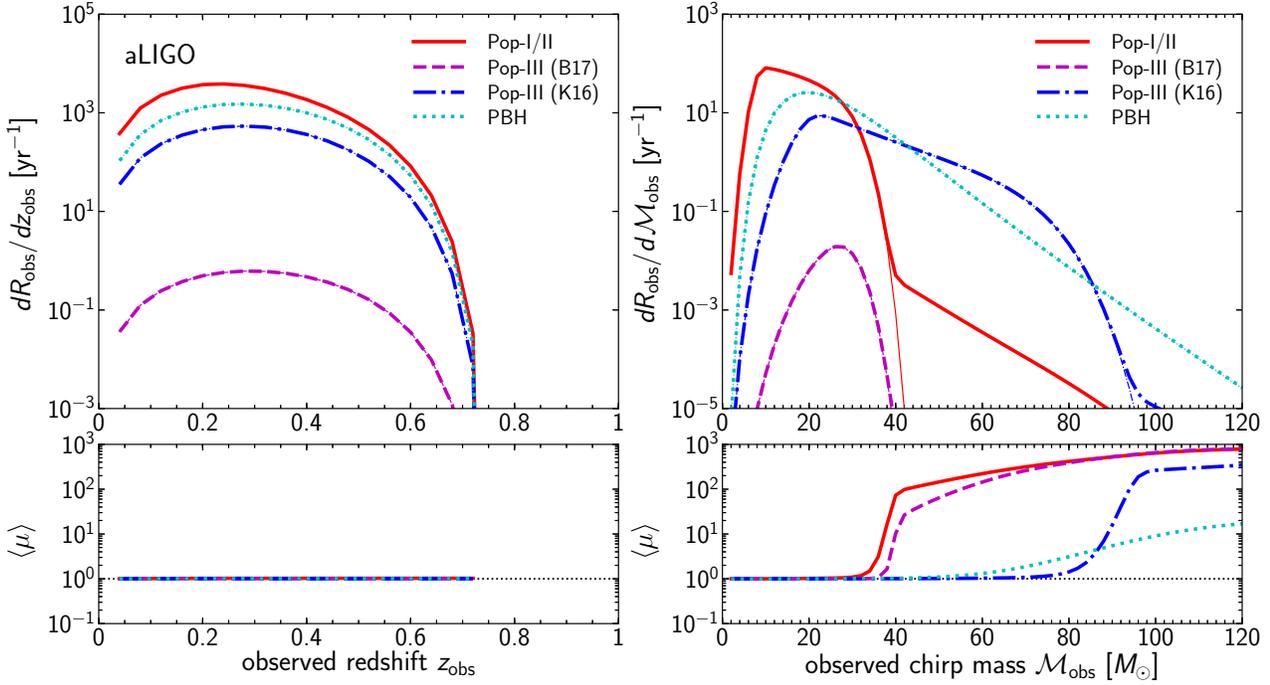}
\end{center}
\caption{The differential distributions of the event rate for
  advanced LIGO. The left panel shows distributions of the observed
  redshift $z_{\rm obs}$ defined in equation~(\ref{eq:z_obs}), whereas
  the right panel shows distributions of the observed chirp mass
  define in equation~(\ref{eq:m_obs}). We plot distributions for the
  four binary BH merger models shown in Figures~\ref{fig:mrate} and
  \ref{fig:dist_bhm}. For each model, thick lines show the
  distributions with the effect of gravitational lensing
  magnification (equation~\ref{eq:robs}) and thin lines show the
  distributions without the lensing effect
  (equation~\ref{eq:robs_nolens}). The mean magnifications are shown
  in the lower panels.
\label{fig:wgobs_aLIGO}}
\end{figure*}

\begin{figure*}
\begin{center}
 \includegraphics[width=0.95\hsize]{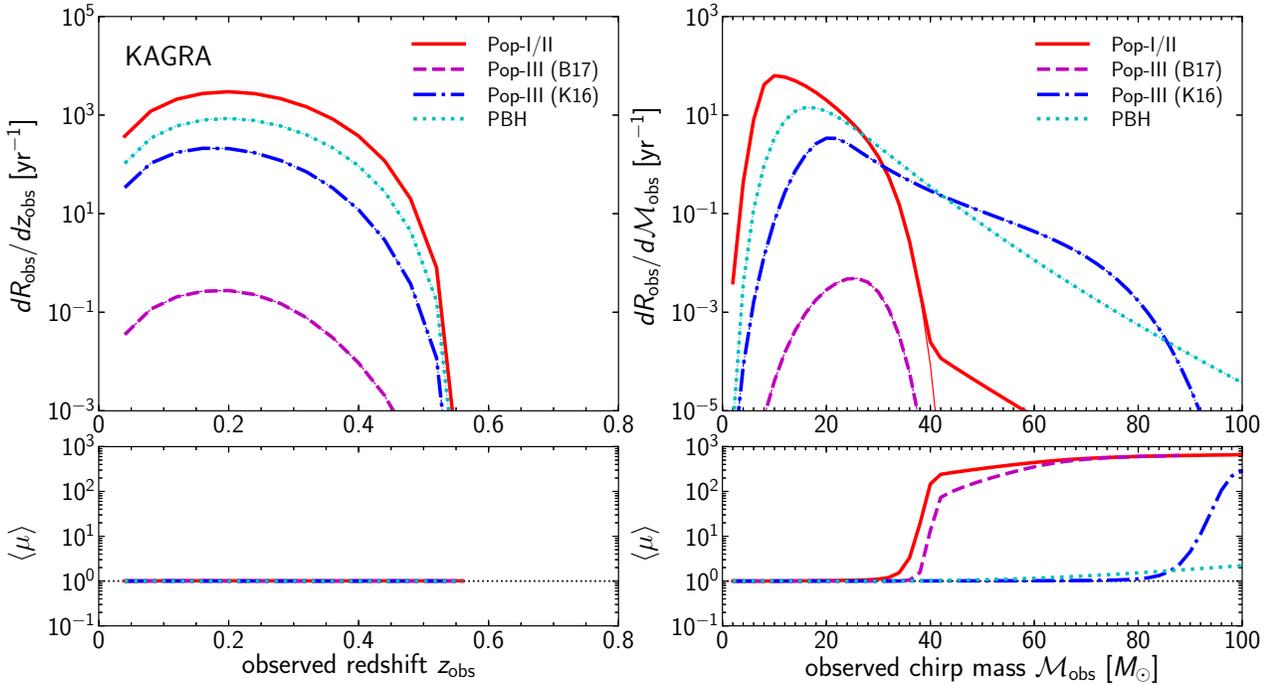}
\end{center}
\caption{Same as Figure~\ref{fig:wgobs_aLIGO}, but for KAGRA.
\label{fig:wgobs_KAGRA}}
\end{figure*}

\begin{figure*}
\begin{center}
 \includegraphics[width=0.95\hsize]{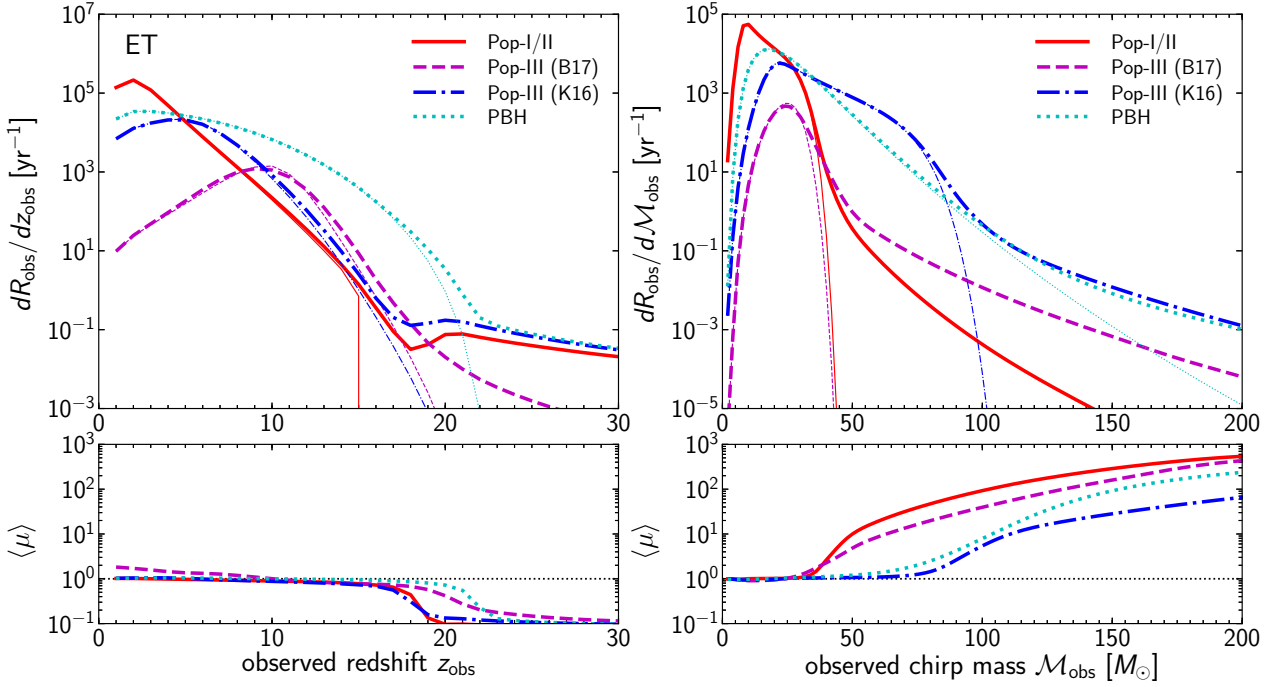}
\end{center}
\caption{Same as Figure~\ref{fig:wgobs_aLIGO}, but for Einstein Telescope.
\label{fig:wgobs_ET}}
\end{figure*}

\begin{figure*}
\begin{center}
 \includegraphics[width=0.95\hsize]{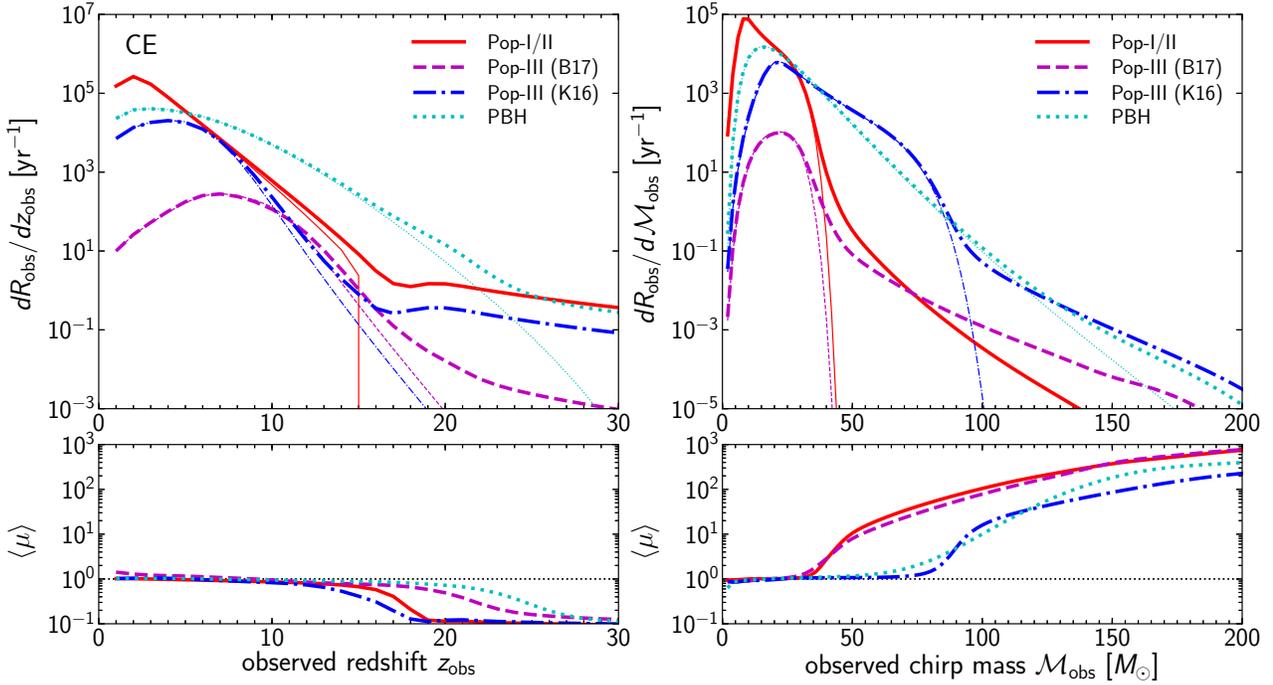}
\end{center}
\caption{Same as Figure~\ref{fig:wgobs_aLIGO}, but for Cosmic Explorer.
\label{fig:wgobs_CE}}
\end{figure*}

\begin{figure*}
\begin{center}
 \includegraphics[width=0.95\hsize]{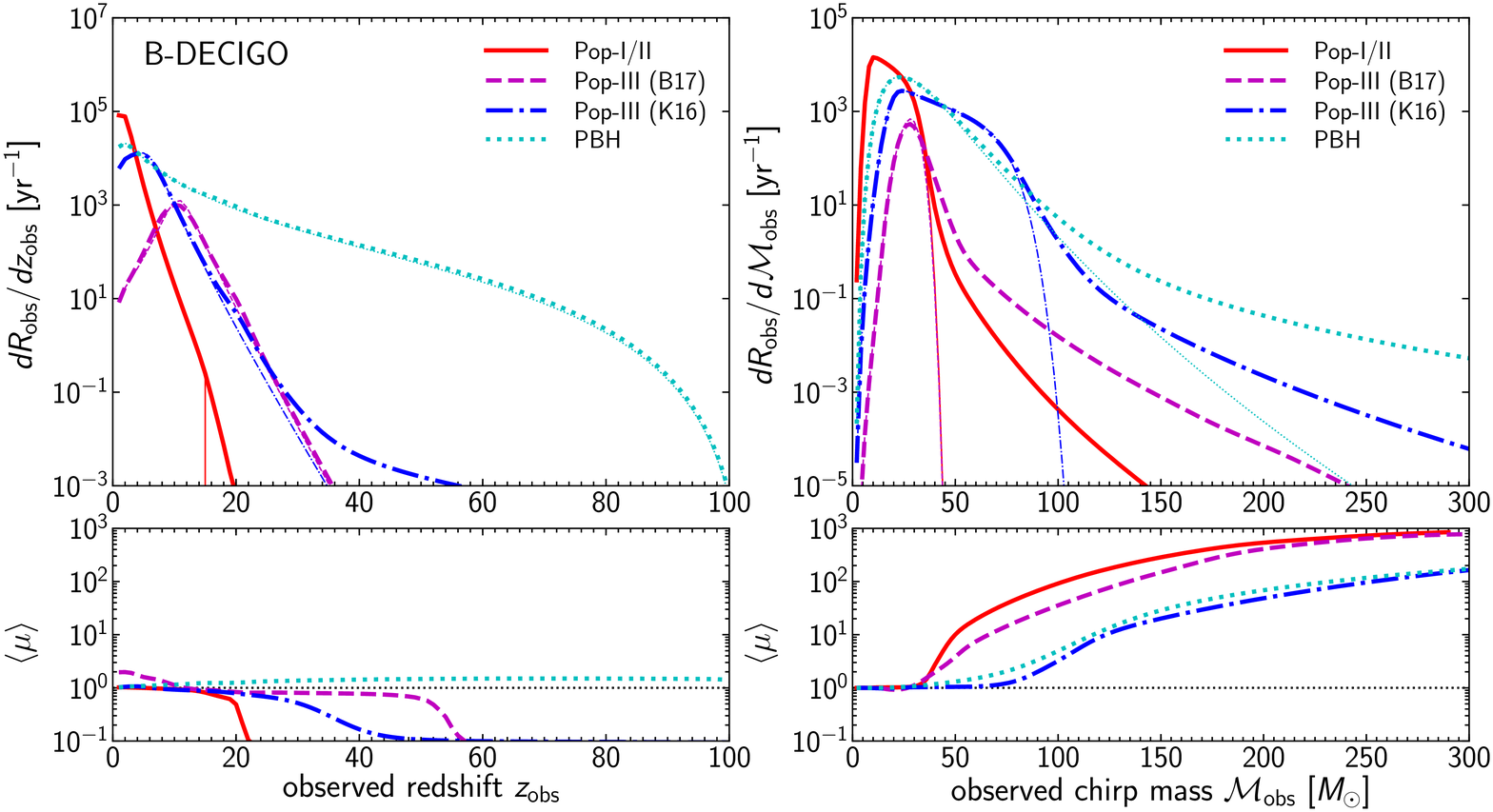}
\end{center}
\caption{Same as Figure~\ref{fig:wgobs_aLIGO}, but for B-DECIGO.
\label{fig:wgobs_preDECIGO}}
\end{figure*}

\section{Results}
\label{sec:results}

\subsection{Distributions in various observatories}
We first derive differential distributions as a function of observed
redshift $z_{\rm obs}$ (equation~\ref{eq:z_obs}) as well as observed
chirp mass $\mathcal{M}_{\rm obs}$ (equation~\ref{eq:m_obs}) for various
gravitational wave observatories summarized in Section~\ref{sec:exps}.
Throughout the paper we adopt the signal-to-noise threshold of
$\rho_{\rm th}=8$ to compute expected distributions.
Figures~\ref{fig:wgobs_aLIGO}, \ref{fig:wgobs_KAGRA},
\ref{fig:wgobs_ET}, \ref{fig:wgobs_CE}, and \ref{fig:wgobs_preDECIGO}
show event rate distributions for advanced LIGO, KAGRA, Einstein
Telescope, Cosmic Explorer, and B-DECIGO, respectively. Here we ignore
the measurement errors and show distributions that would be observed
in absence of any measurement errors. Even without measurement errors,
the event rate distributions are modified due to gravitational lensing
magnification that we cannot be corrected for individual event basis.

We find that the differential distributions are modified due to
gravitational lensing magnification, mainly at high $z_{\rm obs}$ and
high $\mathcal{M}_{\rm obs}$. The high mass tail of the chirp mass
distribution produced by lensing magnification has been discussed in
the literature \citep[e.g.,][]{dai17a,broadhurst18}, which is due to
highly magnified binary BH merger events. To explicitly check this
point in our calculation, we compute the mean magnification as a
function of the observed chirp mass from equation~(\ref{eq:robs}) as
\begin{eqnarray}
\langle\mu\rangle(\mathcal{M}_{\rm obs}) &=& \left(\frac{dR_{\rm
    obs}}{d\mathcal{M}_{\rm obs}}\right)^{-1}\int dz
\int d\mu\,\mu\frac{dP}{d\mu}\frac{d\mathcal{M}}{d\mathcal{M}_{\rm obs}}\nonumber\\
&&\times  \frac{dV}{dz}\frac{R_{\rm
    GW}(z)}{1+z}\frac{dp}{d\mathcal{M}}
S_{\rm lens}(\rho_{\rm th}; \mathcal{M}, z, \mu).
\label{eq:muave}
\end{eqnarray}
The mean magnifications shown in the Figures clearly indicate that the
tail at high $\mathcal{M}_{\rm obs}$ is driven by high magnification
events. 

Furthermore, we find that gravitational lensing magnification produces
a high redshift tail in the $z_{\rm obs}$ distribution. The
distribution of mean magnification computed in a manner 
similar to equation~(\ref{eq:muave}) indicates that this excess is
driven by {\it demagnified} events. Our magnification PDF shown in
Figure~\ref{fig:mag_pdf} suggests that such demagnified ($\mu \sim
0.1$) events are due to strong gravitational lensing. Strong lensing
produces multiple images such that total magnifications of these
multiple images are always larger than unity, but some of the multiple
images can have $\mu<1$. Because of the difficulty in identifying
multiple images in gravitational wave observations, these demagnified
images are also assumed to be observed as distinct events, but due to
lensing demagnifications they have observed redshifts much larger than
their true redshifts, i.e., $z_{\rm obs}>z$. 

\begin{figure}
\begin{center}
 \includegraphics[width=1.0\hsize]{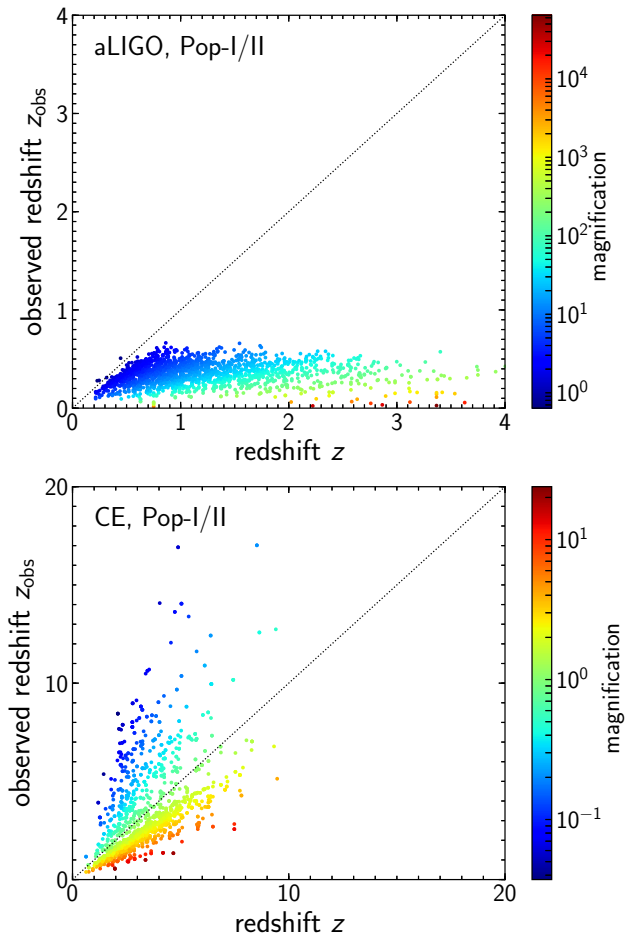}
\end{center}
\caption{The distribution of mock multiple images for advanced
  LIGO ({\it upper}) and Cosmic Explorer ({\it lower}) in the
  $z$-$z_{\rm obs}$ plane, for the case of the Pop-I/II model. We plot
  mock data from 3000 years and 1 year observations for advanced LIGO
  and Cosmic Explorer, respectively.
\label{fig:zcomp}}
\end{figure}

\subsection{Mock strong lens catalogues}
\label{sec:gen_mock}

As shown in the previous Section, high $\mathcal{M}_{\rm obs}$ and
high $z_{\rm obs}$ events can be dominated by very high and low
magnifications, respectively, both of which are produced by strong
lensing (see also Figure~\ref{fig:mag_pdf}). An advantage of our
hybrid approach to compute the magnification PDF is that it also allows
us to explore expected properties of multiple image pairs in
detail, including expected time delays between these multiple images.

To explore the property of multiple images in detail, we construct mock
multiple image catalogues following the methodology developed in
\citet{oguri10b}. We first generate a large mock sample of
gravitational wave events for a given model of the merger rate and
chirp mass distributions. For each mock event, we check whether it is
multiply imaged or not, using the lens model described in
Section~\ref{sec:magpdf_high}. When multiple images are generated, for
each image we compute the signal-to-noise ratio using
equation~(\ref{eq:rho_def}) taking account of gravitational lensing
magnification via equation~(\ref{eq:dis_lum_mag}). For each image, we
randomly assign the parameter $\Theta$ following the PDF of $\Theta$
presented in equation~(\ref{eq:theta_pdf}) to keep the consistency
with the calculation presented in the previous Section. However we
caution that this assumption may be inaccurate, particularly for image
pairs with very short time delays, as the parameter $\Theta$ for these
close pair events should be correlated rather than independent. We
collect events with their signal-to-noise ratio larger than $\rho_{\rm
 th}=8$ to construct a mock catalogue for a given model and observatory.

This mock strong lens catalogue allows us to explore the relation
between various parameters. As a specific example, we check the
distribution of mock strong lens events in the $z$-$z_{\rm obs}$
plane. They are related with each other via equation~(\ref{eq:z_obs}),
which indicates that the deviation from $z=z_{\rm obs}$ is simply
caused by gravitational lensing magnification. Figure~\ref{fig:zcomp}
shows the distributions in the $z$-$z_{\rm obs}$ plane for advanced
LIGO and Cosmic Explorer as examples of second- and third-generation
observatories, respectively. As shown in the Figure, there is a
qualitative difference between the distributions for advanced LIGO and
Cosmic Explorer. In the former case, detectable strong lens events are
dominated by highly magnified events, and hence the observed redshift
$z_{\rm obs}$ is always lower than the true redshift. On the other
hand, in the case of Cosmic Explorer we can observe both magnified and
demagnified events, so that there are events both at
$z>z_{\rm obs}$  and $z<z_{\rm obs}$.

This qualitative difference can be explained by the selection
effect. As shown in Figure~\ref{fig:tau}, the lensing optical depth is
very steep function of redshift out to $z\sim 1$. Because of this
steep growth of the lensing optical depth, strong lens events observed
in the second-generation observatories should be dominated by highly 
magnified high-redshift events \citep[see also][for an extreme
  example]{broadhurst18}. For the specific example shown in  
Figure~\ref{fig:tau}, the median magnification for this strong lens
sample is $\langle \mu\rangle\sim 14$. Such high median magnification
due to the selection effect was also seen in observations of
strongly lensed supernovae. Strongly lensed Type Ia supernovae
recently discovered in relatively shallow surveys, PS1-10afx
\citep{quimby14} and iPTF16geu \citep{goobar17}, have high total
magnifications of $\mu\sim 30-50$ for the same reason as discussed
above \citep[see also][]{quimby14}. In most cases, the
low-magnification counterimages of these high magnification images are
not observable due to their low signal-to-noise ratios.  

In contrast, in the case of Cosmic Explorer we can observe both
magnified and demagnified events. This is because we can detect many
binary BH mergers at $z\sim 2-5$ with Cosmic Explorer without lensing
magnifications (see Figure~\ref{fig:wgobs_CE}), and thanks to the high
sensitive of the observatory, demagnified image of strong lens events at
these redshifts can also be detected easily. These demagnified
multiple images are the origin of apparently very high observed redshift
($z_{\rm obs}\ga 15$) events shown in Figure~\ref{fig:wgobs_CE}.  

\begin{figure*}
\begin{center}
 \includegraphics[width=0.95\hsize]{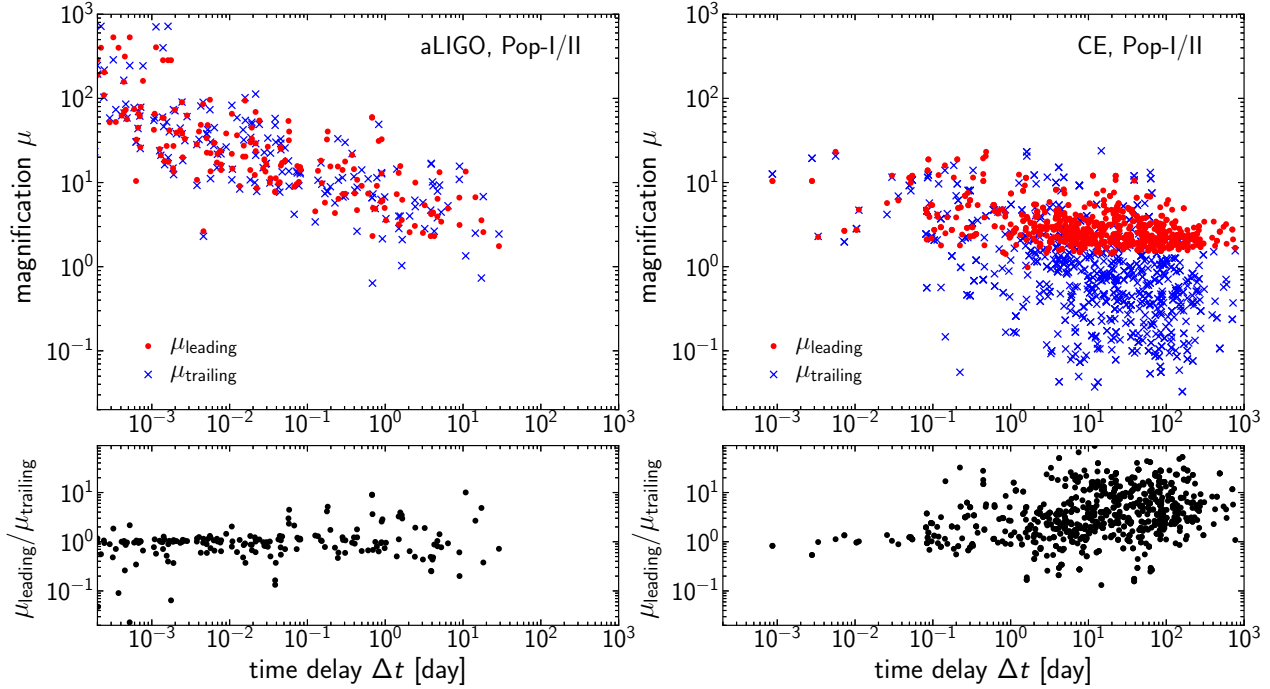}
\end{center}
\caption{Distributions of time delays and magnifications for pairs of
  multiple images from the mock strong lens catalogues. As in
  Figure~\ref{fig:zcomp}, we plot mock data of the Pop-I/II model from
  3000 years and 1 year observations for advanced LIGO ({\it left})
  and Cosmic Explorer ({\it right}), respectively. The upper panels
  show time delays and magnifications of leading ({\it filled
    circles}) and trailing ({\it  crosses}) images for any image pairs
  in the mock catalogues. The bottom panels show time delays and ratios
  of magnifications of leading and trailing images.
\label{fig:tdmag}}
\end{figure*}

\begin{table*}
 \caption{Summary of predicted event rates for various observatories and
   models of binary BH mergers. $R_{\rm obs}$ denotes the total number
   of observed events per year, $R_{\rm sl}$ is the total number of
   strongly lensed events per year, $\langle \mu_{\rm sl}\rangle$ is
   the median magnification of strongly lensed events, $R_{\rm pair}$
   is the total number of observed multiple image pairs per year,
   $\Delta t$ is the median time delay of the observed multiple
   image pairs, and $\langle\mu_{\rm leading}/\mu_{\rm
     trailing}\rangle$ is the median value of the ratio of
   magnifications of leading and trailing images of the observed
   multiple image pairs. Values of $\langle \mu_{\rm sl}\rangle$,
   $R_{\rm pair}$, $\Delta t$, and $\langle\mu_{\rm leading}/\mu_{\rm
     trailing}\rangle$ are derived from the strong lens mock catalogues
   (see Section~\ref{sec:gen_mock}). Values in parentheses for
   $\langle \mu_{\rm sl}\rangle$, $\Delta t$, and $\langle\mu_{\rm
     leading}/\mu_{\rm trailing}\rangle$ denote 68\% ranges, again
   derived from the strong lens mock catalogues. For aLIGO/Pop-III
   (B17) and KAGRA/Pop-III (B17), we fail to construct mock lens
   catalogues because they predict too low strong lens event rates.
\label{tab:rate}}    
 \begin{tabular}{@{}cccccccc}
 \hline
   observatory/model
   & $R_{\rm obs}$ [yr$^{-1}$]
   & $R_{\rm sl}$ [yr$^{-1}$]
   & $\langle\mu_{\rm sl}\rangle$
   & $R_{\rm pair}$ [yr$^{-1}$]
   & $\Delta t$ [day]
   & $\langle\mu_{\rm leading}/\mu_{\rm trailing}\rangle$
   \\
 \hline
    aLIGO/Pop-I/II      & 1.14e+03 & 5.84e$-$01 &  14.35 (3.39--72.71) & 7.77e$-$02 &   0.006 (0.000--0.739) &   1.00 (0.61--1.23) \\ 
    aLIGO/Pop-III (B17) & 2.00e$-$01 & 6.21e$-$05 & --- & --- & --- & --- \\ 
    aLIGO/Pop-III (K16) & 1.68e+02 & 3.89e$-$02 &   6.32 (2.50--27.97) & 3.33e$-$03 &   0.433 (0.013--2.906) &   1.22 (0.82--1.37) \\ 
    aLIGO/PBH           & 4.75e+02 & 1.35e$-$01 &   6.89 (2.40--32.84) & 1.43e$-$02 &   0.124 (0.002--2.853) &   0.92 (0.48--1.54) \\ 
    KAGRA/Pop-I/II      & 6.84e+02 & 1.69e$-$01 &  17.49 (3.30--105.11) & 2.37e$-$02 &   0.002 (0.000--0.090) &   1.00 (0.52--1.19) \\ 
    KAGRA/Pop-III (B17) & 5.58e$-$02 & 3.81e$-$06 & --- & --- & --- & --- \\ 
    KAGRA/Pop-III (K16) & 4.59e+01 & 3.10e$-$03 &   7.65 (2.51--83.11) & 6.67e$-$04 &   0.005 (0.002--0.008) &   1.01 (1.00--1.01) \\ 
    KAGRA/PBH           & 1.93e+02 & 2.00e$-$02 &   7.27 (2.65--45.64) & 3.33e$-$03 &   0.546 (0.139--1.081) &   1.05 (0.81--1.79) \\ 
       ET/Pop-I/II      & 5.54e+05 & 1.12e+03 &   2.10 (0.88--3.55) & 4.56e+02 &  13.741 (1.184--83.138) &   2.36 (0.91--6.75) \\ 
       ET/Pop-III (B17) & 5.96e+03 & 7.38e+01 &   2.41 (1.70--4.32) & 1.50e+01 &  16.518 (0.736--79.897) &   1.95 (0.70--5.10) \\ 
       ET/Pop-III (K16) & 1.13e+05 & 4.86e+02 &   2.10 (0.83--3.40) & 1.74e+02 &  15.094 (1.328--96.548) &   2.61 (0.93--6.91) \\ 
       ET/PBH           & 2.27e+05 & 1.18e+03 &   2.25 (1.36--3.93) & 3.55e+02 &  12.942 (1.042--80.279) &   2.06 (0.80--5.60) \\ 
       CE/Pop-I/II      & 7.31e+05 & 1.60e+03 &   1.88 (0.38--3.09) & 8.36e+02 &  20.600 (2.318--113.044) &   3.64 (1.24--11.20) \\ 
       CE/Pop-III (B17) & 1.54e+03 & 1.51e+01 &   2.44 (1.88--3.98) & 2.60e+00 &   8.266 (0.501--208.184) &   3.02 (1.02--6.55) \\ 
       CE/Pop-III (K16) & 9.96e+04 & 3.96e+02 &   2.07 (0.60--3.64) & 1.82e+02 &  21.283 (1.444--107.229) &   2.90 (0.92--8.78) \\ 
       CE/PBH           & 2.47e+05 & 1.07e+03 &   2.05 (0.71--3.49) & 4.63e+02 &  18.806 (1.290--108.130) &   2.68 (1.01--8.18) \\ 
 B-DECIGO/Pop-I/II      & 2.02e+05 & 4.71e+02 &   2.36 (1.63--4.19) & 9.98e+01 &   8.252 (0.595--56.830) &   1.70 (0.78--4.65) \\ 
 B-DECIGO/Pop-III (B17) & 5.96e+03 & 9.20e+01 &   2.50 (1.76--4.84) & 1.92e+01 &   3.430 (0.188--21.441) &   1.23 (0.50--2.82) \\ 
 B-DECIGO/Pop-III (K16) & 7.66e+04 & 3.86e+02 &   2.27 (1.47--3.94) & 1.22e+02 &  14.577 (1.060--86.073) &   1.88 (0.78--4.78) \\ 
 B-DECIGO/PBH           & 1.31e+05 & 1.41e+03 &   2.63 (1.81--5.43) & 2.70e+02 &   4.965 (0.264--50.640) &   1.29 (0.57--3.29) \\ 
\hline
 \end{tabular}
\end{table*}

\subsection{Time delays between observed multiple image pairs}

One application of the mock strong lens catalogues constructed in
Section~\ref{sec:gen_mock} is that they allow us to estimate
distributions of time delays between multiple image pairs.
From the strong lens mock, we select all pairs of multiple images both
of which are detected, i.e., $\rho>\rho_{\rm th}$. Our mock catalogues
contain strong lens events with more than two (typically four)
images. When more than two images are detected, we consider all the
possible pairs of multiple images and derive time delays between all
these pairs. 

Figure~\ref{fig:tdmag} shows distributions of time delays and
magnifications for all pairs of multiple images in the strong lens mock
catalogues, again for advanced LIGO and Cosmic Explorer as examples of
second- and third-generation observatories, respectively. In addition to
the difference in typical magnifications, we find that typical time
delays are also quite different between advanced LIGO and Cosmic
Explorer.

In the case of advanced LIGO, we preferentially detect binary BH
mergers that are highly magnified by gravitational lensing due to the
selection effect. In most cases, such high magnifications are realized
near the fold or cusp catastrophe, where pairs of multiple images with
similar magnifications are produced. Because they share similar light
paths, time delays between these high magnification image pairs are
very short. We find that time delays for multiple image pairs from
advanced LIGO are indeed short, typically less than a day. Given the
relatively low total event rate of advanced LIGO, it should be
relatively easy to identify strong lensing events from the occurrence
of multiple events in a short time scale. Realistic estimates of the
identifications of such multiple events require to take account of the
effect of the Earth rotation as well as data glitches \citep[see
  also][]{broadhurst18}. We leave the exploration of this for future 
work.

When a source is located very close to the caustic, the wave effect
becomes important. This happens when the time delay between multiple
images near the critical curve is comparable to the wave period 
\citep[e.g.,][]{nakamura99}. Figure~\ref{fig:tdmag} suggests that,
even for images with very high magnifications, $\mu\sim 10^3$, time
delays between multiple image pairs are typically larger than a second
and therefore are much larger than the inverse of frequency of
ground-based gravitational wave observations. Thus the wave effect can
be neglected even for these highly magnified image pairs.

On the other hand, in the case of Cosmic Explorer we can detect many
multiple image pairs from more typical asymmetric image configurations
with large magnification differences. Such asymmetric image
configurations produce image pairs with large time delays. We find
that typical time delays are $10-100$~days, which are much longer than
time delays in the case of advanced LIGO. Together with the high total
event rates, this relatively long time delays make it challenging to  
distinguish such multiple images from two distinct single image
events. Furthermore, long time delays suggest that some of multiple
images cannot be detected in a given observing run, because one of the
multiple images can arrive before or after the observing run. This
``time delay bias'' has been considered in \citet{oguri03} in the
context of gravitationally lensed supernovae, and was also discussed
in \citet{li18}.  

Figure~\ref{fig:tdmag} indicates that there is a clear difference in
the distributions of magnifications between leading and trailing
images of multiple image pairs. In particular, this Figure suggests
that highly demagnified images, which produce a high redshift tail in
the $z_{\rm obs}$ distribution, almost always correspond to trailing
images. This indicates that, when a very high $z_{\rm obs}$ event due
to lensing demagnification is detected, such even should be
accompanied by a much lower $z_{\rm obs}$ event that is observed $\sim
10-100$~days before the high $z_{\rm obs}$ event. Waveforms of
these two multiple image events with high and low $z_{\rm obs}$ should
be similar except for their overall amplitudes. In practice,
demagnified events receive additional frequency dependent phase shift,
which may help identify these strongly lensed multiple image pairs
\citep{dai17b}. 

Table~\ref{tab:rate} summarizes predicted event rates for various
observatories and models of binary BH mergers. Again, this Table
highlights the large difference between ongoing (advanced LIGO
and KAGRA) and future (Einstein Telescope, Cosmic Explorer, and
B-DECIGO) observatories on typical values of magnifications of time
delays. The fractions of strongly lensed events are found to $\sim
10^{-4}$ for advanced LIGO and KAGRA and $\sim 10^{-2}-10^{-3}$ for
Einstein Telescope, Cosmic Explorer, and B-DECIGO. 

\section{Summary}
\label{sec:summary}

In this paper, we have explored the effect of gravitational lensing
magnification on the distribution of binary BH mergers observed by
gravitational wave observatories. For this purpose, we have developed
a hybrid model of the PDF of gravitational lensing magnification, in
which the effects of weak and strong gravitational lensing are
combined. In particular, we derive the magnification PDF by treating
multiple images separately (see Appendix~\ref{sec:magpdf_def}), which
should be appropriate here given the poor angular resolution of
gravitational wave observatories as well as faint electromagnetic
counterparts if at all exist.
 
We have found that pronounced effects of gravitational lensing
magnifications appear at high observed chirp mass
$\mathcal{M}_{\rm obs}$ (equation~\ref{eq:m_obs}) and at high observed
redshift $z_{\rm obs}$ (equation~\ref{eq:z_obs}). The heavy tail of
the distribution at high $\mathcal{M}_{\rm obs}$ is due to highly
magnified strong lens events, which has been recognized in previous
work \citep{dai17a,smith18,broadhurst18}. We have found that highly
{\it demagnified} images of strong lensing events also produce a
heavy tail of the distribution at high $z_{\rm obs}$, which
can be easily detected in future gravitational wave observatories. 
It has been argued that the presence or absence of very high redshift
BH merger events provide an importance clue for discriminating various
binary BH formation models \citep{nakamura16,koushiappas17}, but our
work demonstrates that the effect of gravitational lensing has to be
taken into account carefully in order to properly interpret apparently
very high redshift events. 

Our hybrid approach enables us to explore the expected properties of
strong lensing events detectable in individual gravitational wave
observatories. For instance, in ongoing gravitational wave observatories
such as advanced LIGO and KAGRA, we preferentially observe highly
magnified strong lensing events due to the selection effect. As a
result, we expect to observe pairs of strongly lensed events with
similar magnifications and short time delays of $\la 1$~day,
suggesting that we may be able to identify strongly lensed events from
such image pairs with similar properties. On the other hand, in 
the next generation gravitation wave observatories such as Einstein
Telescope and Cosmic Explorer, strong lensing events are dominated by
those with ``asymmetric'' image configurations with large
magnification ratios and large time delays between multiple
images. Our mock catalogues of strong lens events indicate that highly
demagnified images, which are important source of apparently high
observed redshift events, should be accompanied by a magnified event
that is observed typically $10-100$ days before the demagnified
event. However, the expected long time delays may make it challenging
to identify such pairs of strong lensing events with magnifications
and demagnifications. 

In this paper, we have adopted several simplified assumptions. While we
have assigned the angular orientation function $\Theta$ completely
randomly for different events, values of $\Theta$ should be correlated
for image pairs with short time delays. We have also ignored
measurement errors of observed redshifts and chirp masses when
discussing their distributions. In order to discuss the possibility of
identifying multiple image pairs, we need to take account of the
localization accuracy on the sky as well as the chance probability of
having distinct gravitational wave events with similar waveforms. 
Given the limited amount of information available from binary BH
merger events, it is of great importance to explore the possibility of
identifying multiple images in a realistic situation, in order to
understand the origin of possible extreme events with very high
$\mathcal{M}_{\rm obs}$ or $z_{\rm obs}$ detected in the future.

\section*{Acknowledgements}
I thank T. Broadhurst, L. Dai, M. Sasaki, T. Suyama, H. Tagoshi, and 
M. Takada for useful discussions.
This work was supported in part by World Premier International
Research Center Initiative (WPI Initiative), MEXT, Japan, and JSPS
KAKENHI Grant Number JP18H04572, JP15H05892, and JP18K03693.



\appendix
\section{Magnification PDFs in the presence of multiple images}
\label{sec:magpdf_def}

In this paper, we are interested in the magnification PDF defined in
the source plane, as gravitational wave sources are randomly
distributed in the source plane rather than in the image plane.
In the strong lens regime, the definition of the magnification is
rather ambiguous because a single source produces multiple images.
We argue that the relevant definition of the magnification factor
depends on the observation and the identification scheme of multiple
images. For instance, in survey observations with poor angular
resolutions, such as an imaging survey in the sub-millimetre band
\citep[e.g.,][]{negrello10}, multiple images are not resolved. In this
case, it is appropriate to adopt the total magnification i.e., the sum
of magnification factors of multiple images, as the definition of the
magnification. 

The situation is more complicated when multiple images are resolved. 
In the case of gravitational wave observations, while multiple images
are not resolved spatially given their poor angular resolutions, they
are almost always resolved temporally. However, as discussed in this
paper, the identification of multiple images is not straightforward in
gravitational wave observations again given their poor spatial
resolutions. Thus in this paper we treat multiple images separately. 
In this case, the number of sources is not conserved by strong
gravitational lensing i.e., the number of sources in the image plane
differs from the number of sources in the source plane. 

In practice, from the strong lens mock sample at a give source
redshift constructed in Section~\ref{sec:magpdf_high}, we derive two
types of magnification PDFs. First, we derive the magnification PDF in
which multiple images are grouped together. For $j$-th strong lens
system, we compute the total magnification as
\begin{equation}
\mu_{j, {\rm tot}} = \sum_k \mu_{j,k},
\end{equation}
where the summation $k$ runs over multiple images of the $j$-th mock
strong lens system, and $\mu_{j,k}$ denotes the magnification of the
individual image for the $j$-th mock strong lens system. We then
compute the magnification PDF in the $i$-th magnification bin with
$\mu_{i,{\rm min}}<\mu_i<\mu_{i,{\rm max}}$ as 
\begin{equation}
\frac{dP_{\rm sl, lens}}{d\mu_i}=\frac{1}{N_{\rm s}}\sum_j
\Theta(\mu_{j,{\rm tot}}-\mu_{i,{\rm min}})\Theta(\mu_{i,{\rm max}}-\mu_{j,{\rm tot}}),
\end{equation}
where $\Theta$ is the Heaviside step function, $N_s$ is the total
number of sources at this source redshift that are randomly
distributed to generate the mock lens sample, and the summation runs
over the mock lens sample. This magnification PDF is normalized such
that the integral over the magnification gives the lensing optical
depth $\tau_{\rm multi}$ i.e., the total probability of strong
gravitational lensing with multiple images 
\begin{equation}
\int d\mu \frac{dP_{\rm sl, lens}}{d\mu_i} = \tau_{\rm multi}.
\end{equation}
Next, we define another magnification PDF for which multiple images of
individual mock lens system are treated separately. Using the
notations defined above, we can derive this magnification PDF as
\begin{equation}
\frac{dP_{\rm sl, img}}{d\mu_i}=\frac{1}{N_{\rm s}}\sum_j\sum_k
\Theta(\mu_{j,k}-\mu_{i,{\rm min}})\Theta(\mu_{i,{\rm max}}-\mu_{j,k}).
\end{equation}
This magnification PDF has a different normalization from $dP_{\rm sl,
  lens}/d\mu$. Specifically, $dP_{\rm sl, img}/d\mu$ is normalized
such that
\begin{equation}
\int d\mu \frac{dP_{\rm sl, img}}{d\mu_i} = \langle N_{\rm img}\rangle
\tau_{\rm multi},
\label{eq:tau_multi_n}
\end{equation}
where $\langle N_{\rm img}\rangle$ is the average number of multiple
images for the mock strong lens systems. Since most strong lens
systems have two or four multiple images, we expect $2<\langle N_{\rm
  img}\rangle<4$. 

As shown in Section~\ref{sec:magpdf_comb}, we combine the magnification
PDF from the strong lens mock sample with that at low magnification
derived in Section~\ref{sec:magpdf_low} to obtain the total
magnification PDF. However, if we simply sum up these magnification
PDFs, the resulting magnification PDF does not satisfy the correct
normalization condition. Therefore we tweak the normalization of the
magnification PDF at low magnification to ensure the normalization
\begin{equation}
\frac{dP}{d\mu}=\frac{1-\tau_{\rm multi}}{\int d\mu\,dP_{\rm wl}/d\mu}
\frac{dP_{\rm wl}}{d\mu}+\frac{dP_{\rm sl}}{d\mu}.
\end{equation}
We note that the prefactor $(1-\tau_{\rm multi})/\int d\mu\,dP_{\rm
  wl}/d\mu$ is in fact quite close to unity with a typical deviation
of $\sim 3$\% or so, indicating that this correction is a minor
correction. When we adopt $dP_{\rm sl}/d\mu=dP_{\rm sl, lens}/d\mu$,
we can easily show that  
\begin{equation}
\int d\mu\frac{dP}{d\mu}=1,
\end{equation}
which is expected from the conservation of the number of sources by
gravitational lensing. In contrast, when we adopt $dP_{\rm
  sl}/d\mu=dP_{\rm sl, img}/d\mu$, the number of sources no longer
conserves as strong lensing, once multiple images are treated
separately, increases the number of observed events.  In this case, 
the normalization exceeds unity
\begin{equation}
\int d\mu\frac{dP}{d\mu}=1+\tau_{\rm multi}(\langle N_{\rm img}\rangle-1)>1,
\end{equation}
although we note that $\tau_{\rm multi}\ll 1$ and hence the excess is
small. As mentioned above, in this paper we adopt $dP_{\rm
  sl}/d\mu=dP_{\rm sl, img}/d\mu$ and treat multiple images
separately. 

The magnification PDF constructed above does not guarantees $\langle
\mu\rangle=1$ which is expected for the source plane magnification PDF
\citep[e.g.,][]{takahashi11}, even when we adopt the total
magnification, $dP_{\rm sl}/d\mu=dP_{\rm sl, lens}/d\mu$. To correct
for this, for each source redshift bin we compute the magnification
shift parameter
\begin{equation}
\mu_{\rm shift}=1-\int d\mu\frac{dP}{d\mu}\mu,
\end{equation}
where $dP/d\mu$ is computed using $dP_{\rm sl}/d\mu=dP_{\rm sl,
  lens}/d\mu$, and uniformly shift the magnification in the
magnification PDF as 
\begin{equation}
\mu \rightarrow \mu+\mu_{\rm shift},
\end{equation}
which ensures $\langle \mu\rangle=1$. We apply this shift of the
magnification even when we adopt $dP_{\rm sl}/d\mu=dP_{\rm sl,
  img}/d\mu$ that is used in our main analysis. Again, this shift is
quite minor with $|\mu_{\rm shift}| \la 0.02$ for $z_{\rm s}\la 10$
and $|\mu_{\rm shift}| \la 0.1$ for $z_{\rm s}\la 100$. 

\label{lastpage}

\end{document}